\documentclass{article}

\usepackage{latexsym}
\usepackage{graphicx}

\def\gsize{0.35} %グラフのサイズ　<0.74では線が細くなってしまう
\def\gs2{0.60} %グラフのサイズ　<0.74では線が細くなってしまう

\title{Analysis of on-line learning \\
when a moving teacher goes around a true teacher}

\author{Seiji MIYOSHI$^*$ and Masato OKADA$^\dagger$\\
${}^{*}$Department of Electronic Engineering, 
  Kobe City College of Technology, \\
  8--3 Gakuenhigashimachi, Nishi-ku, 
  Kobe-shi, 651--2194 Japan\\
${}^{\dagger}$Division of Transdisciplinary Sciences, 
  Graduate School of Frontier Sciences, \\
 The University of Tokyo, 
  5--1--5 Kashiwanoha, Kashiwa-shi, Chiba, 277--8561 Japan\\
 RIKEN Brain Science Institute,
  2--1 Hirosawa, Wako-shi, Saitama, 351--0198 Japan\\
 JST PRESTO}

\begin{document}

\maketitle

\begin{abstract}
In the framework of on-line learning,
a learning machine might move around a teacher
due to the differences in structures or output functions
between the teacher and the learning machine or 
due to noises.
The generalization performance 
of a new student supervised by a moving 
machine has been analyzed.
A model composed of a true teacher,
a moving teacher and a student 
that are all linear perceptrons with noises
has been treated analytically
using statistical mechanics. 
It has been proven that
the generalization errors of a student
can be smaller than
that of a moving teacher,
even if the student only uses examples 
from the moving teacher.
\end{abstract}


%**********************************************************
\section{Introduction}
%**********************************************************
Learning is to infer the underlying rules that dominate 
data generation using observed data.
The observed data are input-output pairs from a teacher.
They are called examples.
Learning can be roughly classified into batch learning and
on-line learning \cite{Saad}.
In batch learning, some given examples are used repeatedly.
In this paradigm, a student becomes to give correct answers
after training if that student has an adequate degree of freedom.
However, it is necessary to have a long amount of time and 
a large memory in which many examples
may be stored.
On the contrary, examples used once are discarded
in on-line learning.
In this case, a student cannot give correct answers 
for all examples used in training.
However, there are some merits, for example,
a large memory for storing many examples isn't necessary
and 
it is possible to follow a time variant teacher. 

Recently, we \cite{Hara,PRE} 
have analyzed the generalization performance
of ensemble learning
\cite{Abe,www.boosting.org,Krogh,Urbanczik}
in a framework of on-line learning
using a statistical mechanical method \cite{Saad,NishimoriE}.
In that process, the following points
are proven subsidiarily.
The generalization error doesn't approach zero
when the student is a simple perceptron and the teacher
is a committee machine \cite{IBIS2004} 
or a non-monotonic perceptron \cite{NC200503}.
Therefore, models like these can be called 
unlearnable cases \cite{Inoue,Inoue2}.
The behavior of a student in an unlearnable case
depends on the learning rule.
That is, the student vector asymptotically
converges in one direction using Hebbian learning.
On the contrary, the student vector 
doesn't converge in one direction but continues moving
using perceptron learning or AdaTron learning.
In the case of a non-monotonic teacher,
the student's behavior can be expressed by continuing to 
go around the teacher, 
keeping a constant direction cosine with the teacher.

Considering the applications of statistical learning theories,
investigating the 
system behaviors of unlearnable cases
is very significant
since real world problems seem to 
include many unlearnable cases.
In addition, 
a learning machine may continue going around a teacher
in the unlearnable cases as mentioned above.
Here, let us consider a new student that
is supervised by a moving learning machine.
That is, we consider a student that
uses the input-output pairs of 
a moving teacher as training examples
and we investigate the generalization performance 
of a student with a true teacher.
Note that the examples used by the student
are only from the moving teacher 
and the student can't directly observe the outputs
of the true teacher.
In a real human society, a teacher
that can be observed by a student
doesn't always present the correct answer.
In many cases, the teacher is learning
and continues to vary.
Therefore, the analysis of such a model
is interesting for considering the 
analogies between statistical learning theories 
and a real society.

In this paper, we treat a model 
in which a true teacher, a moving teacher and a student
are all linear perceptrons \cite{Hara} with noises,
as the simplest model in which 
a moving teacher continues going around a true teacher.
We calculate the order parameters and the generalization errors
analytically using a statistical mechanical method 
in the framework of on-line learning.
As a result, it is proven that
a student's generalization errors
can be smaller than that of the moving teacher.
That means the student can be cleverer than the moving teacher
even though the student uses only the examples of the moving teacher.

%**********************************************************
\section{Model}
%**********************************************************
Three linear perceptrons are treated in this paper:
a true teacher, a moving teacher and a student.
Their connection weights are
$\mbox{\boldmath $A$},\mbox{\boldmath $B$}$ and
$\mbox{\boldmath $J$}$, respectively.
For simplicity, the connection weight of the true teacher,
that of the moving teacher and that of the student
are simply called the true teacher, the moving teacher, and
the student, respectively.
The true teacher 
$\mbox{\boldmath $A$}=(A_1,\ldots,A_N)$,
the moving teacher
$\mbox{\boldmath $B$}=(B_1,\ldots,B_N)$,
the student
$\mbox{\boldmath $J$}=(J_1,\ldots,J_N)$, and
input $\mbox{\boldmath $x$}=(x_1,\ldots,x_N)$
are $N$ dimensional vectors.
Each component $A_i$ of $\mbox{\boldmath $A$}$
is drawn from ${\cal N}(0,1)$ independently and fixed,
where ${\cal N}(0,1)$ denotes the Gaussian distribution with
a mean of zero and a variance unity.
Each of the components $B_i^0, J_i^0$
of the initial values of 
$\mbox{\boldmath $B$},\mbox{\boldmath $J$}$
are drawn from ${\cal N}(0,1)$ independently.
Each component $x_i$ of $\mbox{\boldmath $x$}$
is drawn from ${\cal N}(0,1/N)$ independently.
Thus,
\begin{eqnarray}
\left\langle A_i\right\rangle &=& 0, \ 
\left\langle (A_i)^2\right\rangle=1, \\
\left\langle B_i^0\right\rangle &=& 0, \ 
\left\langle (B_i^0)^2\right\rangle=1,\\
\left\langle J_i^0\right\rangle &=& 0, \ 
\left\langle (J_i^0)^2\right\rangle=1,\\
\left\langle x_i\right\rangle &=& 0, \ 
\left\langle (x_i)^2\right\rangle=\frac{1}{N},
\end{eqnarray}
where $\langle \cdot \rangle$ denotes a mean.

In this paper, the thermodynamic limit $N\rightarrow \infty$
is also treated. Therefore,
\begin{equation}
\|\mbox{\boldmath $A$}\|=\sqrt{N},\ \ 
\|\mbox{\boldmath $B$}^0\|=\sqrt{N},\ \ 
\|\mbox{\boldmath $J$}^0\|=\sqrt{N},\ \ 
\|\mbox{\boldmath $x$}\|=1,
\label{eqn:xBJ}
\end{equation}
where $\|\cdot \|$ denotes a vector norm.
Generally, norms $\|\mbox{\boldmath $B$}\|$ and $\|\mbox{\boldmath $J$}\|$
of the moving teacher and the student
change as the time step proceeds.
Therefore, the ratios $l_B$ and $l_J$ of the norms to $\sqrt{N}$
are introduced and are called the length of the moving teacher
and the length of the student. That is,
$\|\mbox{\boldmath $B$}\|=l_B\sqrt{N}$，
$\|\mbox{\boldmath $J$}\|=l_J\sqrt{N}$.

The outputs of the true teacher, the moving teacher, 
and the student are
$y^m+n_{A}^m$, 
$v^m l_B^m+n_B^m$, and
$u^m l_J^m+n_J^m$, respectively.
Here, 
\begin{eqnarray}
y^m&=&
 \mbox{\boldmath $A$}\cdot \mbox{\boldmath $x$}^m, \label{eqn:hatv}\\
v^m l_B^m &=& 
 \mbox{\boldmath $B$}^m\cdot \mbox{\boldmath $x$}^m, \label{eqn:v}\\
u^m l_J^m &=& 
 \mbox{\boldmath $J$}^m\cdot \mbox{\boldmath $x$}^m, \label{eqn:u}
\end{eqnarray}
and
\begin{eqnarray}
n_{A}^m&\sim&{\cal N}(0,\sigma_{A}^2),\\
n_B^m&\sim&{\cal N}(0,\sigma_B^2),\\
n_J^m&\sim&{\cal N}(0,\sigma_J^2).
\end{eqnarray}
where $m$ denotes the time step.
That is,
the outputs of the true teacher, the moving teacher and the student
include independent Gaussian noises with variances of 
$\sigma_{A}^2,\sigma_B^2$, and $\sigma_J^2$, respectively.
Then, the
$y^m$, $v^m$, and $u^m$
of Eqs.(\ref{eqn:hatv})--(\ref{eqn:u})
obey the Gaussian distributions with a mean of zero and 
a variance unity.

In the model treated in this paper,
the moving teacher $\mbox{\boldmath $B$}$
is updated using an input $\mbox{\boldmath $x$}$
and an output of the true teacher $\mbox{\boldmath $A$}$
for the input $\mbox{\boldmath $x$}$.
The student $\mbox{\boldmath $J$}$
is updated by using an input $\mbox{\boldmath $x$}$
and an output of the moving teacher $\mbox{\boldmath $B$}$
for the input $\mbox{\boldmath $x$}$.
Let us define an error between the true teacher and 
the moving teacher by 
the squared error of their outputs.
That is,
\begin{equation}
\epsilon_B^m \equiv 
\frac{1}{2}\left( y^m+n_{A}^m - v^ml_B^m -n_B^m\right)^2.
\label{eqn:hate}
\end{equation}

The moving teacher is considered to
use the gradient method for learning.
That is,
\begin{eqnarray}
\mbox{\boldmath $B$}^{m+1}
&=& \mbox{\boldmath $B$}^{m} 
   -\eta_B 
    \frac{\partial \epsilon_B^m}{\partial \mbox{\boldmath $B$}^{m}}\\
&=& \mbox{\boldmath $B$}^{m} 
   +\eta_B \left( y^m+n_{A}^m - v^ml_B^m -n_B^m\right)
   \mbox{\boldmath $x$}^{m},
\label{eqn:Bupdate}
\end{eqnarray}
where, $\eta_B$ denotes the learning rate
of the moving teacher and is a constant number.

In the same manner, 
let us define an error between the moving teacher and 
the student by 
the squared error of their outputs.
That is,
\begin{equation}
\epsilon_{BJ}^m \equiv 
\frac{1}{2}\left( v^ml_B^m+n_B^m-u^ml_J^m-n_J^m\right)^2.
\label{eqn:e}
\end{equation}

The student is considered to
use the gradient method for learning.
That is,
\begin{eqnarray}
\mbox{\boldmath $J$}^{m+1}
&=& \mbox{\boldmath $J$}^{m} 
   -\eta_J \frac{\partial \epsilon_{BJ}^m}{\partial \mbox{\boldmath $J$}^{m}}\\
&=& \mbox{\boldmath $J$}^{m} 
   +\eta_J \left( v^ml_B^m+n_B^m-u^ml_J^m-n_J^m\right)
   \mbox{\boldmath $x$}^{m},
\label{eqn:Jupdate}
\end{eqnarray}
where, $\eta_J$ denotes a learning rate
of the student and is a constant number.

Generalizing the learning rules, 
Eqs.(\ref{eqn:Bupdate}) and (\ref{eqn:Jupdate})
can be expressed as
\begin{eqnarray}
\mbox{\boldmath $B$}^{m+1}&=& \mbox{\boldmath $B$}^{m}
 +g\left( y^m+n_{A}^m,v^ml_B^m+n_B^m\right)
 \mbox{\boldmath $x$}^{m}, \label{eqn:B} \\
\mbox{\boldmath $J$}^{m+1}&=&\mbox{\boldmath $J$}^{m}
 +f\left( v^ml_B^m+n_B^m,u^ml_J^m+n_J^m\right)
 \mbox{\boldmath $x$}^{m} \label{eqn:J},
\end{eqnarray}
respectively.

Let us define an error between the true teacher and 
the student by 
the squared error of their outputs.
That is,
\begin{equation}
\epsilon_J^m \equiv 
\frac{1}{2}\left( y^m+n_{A}^m -u^ml_J^m-n_J^m\right)^2.
\label{eqn:bare}
\end{equation}

%**********************************************************
\section{Theory}
%**********************************************************
%++++++++++++++++++++++++++++++++++++++++++++++++++++++++++
\subsection{Generalization Error}
%++++++++++++++++++++++++++++++++++++++++++++++++++++++++++
One purpose of a statistical learning theory
is to theoretically obtain generalization errors.
Since a generalization error is the mean of errors 
for the true teacher 
over the distribution of the new input and noises,
the generalization error $\epsilon_{Bg}$ of the moving teacher
and $\epsilon_{Jg}$ of the student
are calculated as follows.
The superscripts $m$, which represent the time steps, are omitted 
for simplicity.
\begin{eqnarray}
\epsilon_{Bg}
&=& \int d\mbox{\boldmath $x$} dn_{A} dn_B 
         P(\mbox{\boldmath $x$}, n_{A}, n_B)
         \epsilon_B
         \\
&=& \int dy dv  dn_{A} dn_B 
    P(y, v, n_{A}, n_B) \nonumber \\
& & \times \frac{1}{2}
    \left( y+n_{A} - vl_B -n_B\right)^2 \\
&=& \frac{1}{2}
    \left(
    -2R_B l_B + (l_B)^2 + 1 + \sigma_{A}^2 + \sigma_B^2
    \right),
    \label{eqn:hateg} \\
\epsilon_{Jg} 
&=& \int d\mbox{\boldmath $x$} dn_{A} dn_J 
         P(\mbox{\boldmath $x$}, n_{A}, n_J)
         \epsilon_J
         \\
&=& \int dy du dn_{A} dn_J 
    P(y, u, n_{A}, n_J) \nonumber \\
& & \times \frac{1}{2}
    \left(y+n_{A} - ul_J - n_J \right)^2 \\
&=& \frac{1}{2}
    \left(
    -2R_J l_J + (l_J)^2 + 1 + \sigma_{A}^2 + \sigma_J^2
    \right).
    \label{eqn:bareg}
\end{eqnarray}

In addition, let us calculate the mean $\epsilon_{BJg}$ of the error between 
the student and the moving teacher as follows:
\begin{eqnarray}
\epsilon_{BJg}
&=& \int d\mbox{\boldmath $x$} dn_B dn_J 
         P(\mbox{\boldmath $x$}, n_B, n_J)
         \epsilon_{BJ}
         \\
&=& \int dv du dn_B dn_J P(v, u, n_B, n_J) \nonumber \\
& & \times \frac{1}{2}
    \left( vl_B+n_B - ul_J - n_J\right)^2 \\
&=& \frac{1}{2}
    \left(
    -2R_{BJ} l_B l_J + (l_J)^2 + (l_B)^2 + \sigma_B^2 + \sigma_J^2
    \right).
    \label{eqn:eg}
\end{eqnarray}

\begin{figure}[htbp]
\begin{center}
\includegraphics[width=\gsize\linewidth,keepaspectratio]{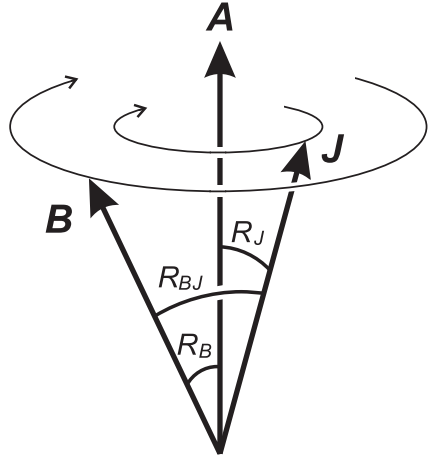}
%-----------------------------------------------
\caption{True teacher $\mbox{\boldmath $A$}$,
moving teacher $\mbox{\boldmath $B$}$ and 
student $\mbox{\boldmath $J$}$.
$R_B, R_J$, and $R_{BJ}$ are direction cosines.}
%-----------------------------------------------
\label{fig:hatBBJ}
\end{center}
\end{figure}

Here, the integration has been executed 
using the following:
$y$, $v$ and $u$ obeys ${\cal N}(0,1)$.
The covariance between $y$ and $v$ is $R_B$,
between $y$ and $u$ is $R_J$,
and between $v$ and $u$ is $R_{BJ}$,
where
\begin{eqnarray}
R_B &\equiv& \frac{\mbox{\boldmath $A$}\cdot \mbox{\boldmath $B$}} 
 {\|\mbox{\boldmath $A$}\|\|\mbox{\boldmath $B$}\|},\ \ \ 
R_J \equiv \frac{\mbox{\boldmath $A$}\cdot \mbox{\boldmath $J$}}
 {\|\mbox{\boldmath $A$}\|\|\mbox{\boldmath $J$}\|},\ \ \ 
R_{BJ} \equiv \frac{\mbox{\boldmath $B$}\cdot \mbox{\boldmath $J$}}
 {\|\mbox{\boldmath $B$}\|\|\mbox{\boldmath $J$}\|}. \label{eqn:R}
\end{eqnarray}
Eq.(\ref{eqn:R}) means that 
$R_B$, $R_J$, and $R_{BJ}$ are direction cosines.
$n_{A}$, $n_B$, and $n_J$
are all independent with other probabilistic variables.
The true teacher $\mbox{\boldmath $A$}$,
the moving teacher $\mbox{\boldmath $B$}$,
the student $\mbox{\boldmath $J$}$,
and the relationship among $R_B, R_J$, and $R_{BJ}$
are shown in Fig.\ref{fig:hatBBJ}.

%++++++++++++++++++++++++++++++++++++++++++++++++++++++++++
\subsection{Differential equations of order parameters and
their analytical solutions}
%++++++++++++++++++++++++++++++++++++++++++++++++++++++++++
To make analysis easy, the following auxiliary order parameters
are introduced:
\begin{eqnarray}
r_B&\equiv&R_B l_B, \label{eqn:hatrdef} \\
r_J&\equiv&R_J l_J, \label{eqn:barrdef} \\
r_{BJ}&\equiv&R_{BJ} l_B l_J.         \label{eqn:rdef}
%L_B&\equiv&l_B^2,           \label{eqn:LBdef}  \\
%L_J&\equiv&l_J^2.           \label{eqn:LJdef}  
\end{eqnarray}

Simultaneous differential equations
in deterministic forms \cite{NishimoriE} have been obtained
that describe
the dynamical behaviors of order parameters
based on self-averaging
in the thermodynamic limits as follows:
\begin{eqnarray}
\frac{dr_B}{dt}&=& \langle gy\rangle, \label{eqn:dhatrdt} \\
\frac{dr_J}{dt}&=& \langle fy\rangle, \label{eqn:dbarrdt}\\
\frac{dr_{BJ}}{dt}&=&
l_J \langle gu\rangle + l_B \langle fv\rangle + \langle gf \rangle, 
\label{eqn:drdt}\\
\frac{dl_B}{dt}&=&
\langle gv \rangle +\frac{\langle g^2 \rangle}{2l_B}, \label{eqn:dlBdt3} \\
\frac{dl_J}{dt}&=&
\langle fu \rangle +\frac{\langle f^2 \rangle}{2l_J}. \label{eqn:dlJdt3}
\end{eqnarray}

Since linear perceptrons are treated in this paper,
the sample averages that appeared in the above 
equations can be calculated easily as follows:
\begin{eqnarray}
\langle gu\rangle &=& \eta_B(r_J-r_{BJ})/l_J, \label{eqn:gu} \\
\langle fv\rangle &=& \eta_J(l_B-r_{BJ}/l_B), \label{eqn:fv} \\
\langle gf \rangle &=& 
\eta_B \eta_J (r_B-r_J-l_B^2+r_{BJ}-\sigma_B^2), \label{eqn:gf} \\
\langle fy\rangle&=& \eta_J (r_B-r_J), \label{eqn:fhatv} \\
\langle gy\rangle&=& \eta_B (1-r_B), \label{eqn:ghatv} \\
\langle gv\rangle&=&\eta_B (r_B/l_B-l_B), \label{eqn:gv} \\
\langle g^2\rangle&=&
\eta_B^2(1+\sigma_{A}^2+\sigma_B^2+l_B^2-2r_B), \label{eqn:g2} \\
\langle fu\rangle&=&\eta_J(r_{BJ}/l_J-l_J), \label{eqn:fu} \\
\langle f^2\rangle&=&
\eta_J^2(l_B^2+l_J^2+\sigma_B^2+\sigma_J^2-2r_{BJ}). \label{eqn:f2}
\end{eqnarray}

Since each components of the true teacher $\mbox{\boldmath $A$}$,
the initial value of the moving teacher $\mbox{\boldmath $B$}$,
and the initial value of the student $\mbox{\boldmath $J$}$
are drawn from ${\cal N}(0,1)$ independently
and because the thermodynamic limit $N\rightarrow \infty$
is also treated,
they are all orthogonal to each other in the initial state.
That is,
\begin{equation}
R_B^0=R_J^0=R_{BJ}^0=0.
\label{eqn:Rinit}
\end{equation}

In addition,
\begin{equation}
l_B^0=l_J^0=1.
\label{eqn:linit}
\end{equation}

By using Eqs.(\ref{eqn:gu})--(\ref{eqn:linit}),
the simultaneous differential equations 
Eqs.(\ref{eqn:dhatrdt})--(\ref{eqn:dlJdt3})
can be solved analytically as follows:
\begin{eqnarray}
r_B&=&1-e^{-\eta_B t}, \label{eqn:hatr} \\
r_J &=& 
   1+\frac{\eta_B}{\eta_J-\eta_B}e^{-\eta_J t}
   -\frac{\eta_J}{\eta_J-\eta_B}e^{-\eta_B t},
   \label{eqn:barr} \\
r_{BJ}&=&
     - \frac{D}{\eta_B\eta_J-\eta_B-\eta_J} \nonumber \\
  &+& \frac{2\eta_J-\eta_B}{\eta_B-\eta_J}e^{-\eta_B t}
      + \frac{\eta_B}{\eta_J-\eta_B}e^{-\eta_J t} \nonumber \\
  &+& \frac{\eta_J}{\eta_J-\eta_B}Ce^{\eta_B(\eta_B-2)t}
     +Ee^{(\eta_B\eta_J-\eta_B-\eta_J)t}, \label{eqn:r} \\
%L_B &=& 3-C-2e^{-\eta_B t}+Ce^{\eta_B(\eta_B-2)t}, \label{eqn:LB} \\
l_B^2 &=& 3-C-2e^{-\eta_B t}+Ce^{\eta_B(\eta_B-2)t}, \label{eqn:LB} \\
%L_J &=&
l_J^2 &=&
   - \frac{G}{\eta_J(\eta_J-2)} \nonumber \\
   &+& \frac{F}{\eta_B(\eta_B-2)-\eta_J(\eta_J-2)}e^{\eta_B(\eta_B-2) t} 
       \nonumber \\
   &+& \frac{2\eta_B}{\eta_J-\eta_B}e^{-\eta_J t}
       -\frac{2\eta_J}{\eta_J-\eta_B}e^{-\eta_B t} \nonumber \\
   &-&  \frac{2\eta_J E}{\eta_B-\eta_J}e^{(\eta_B\eta_J-\eta_B-\eta_J)t}
       +He^{\eta_J(\eta_J-2) t}, \label{eqn:LJ}
\end{eqnarray}
where
\begin{eqnarray}
C &=& 2-\frac{\eta_B}{2-\eta_B}(\sigma_A^2+\sigma_B^2), \\
D &=& \eta_B(1-\eta_J \sigma_B^2)+\eta_J(1-\eta_B)\left(3-C\right), \\
E &=&
   \frac{-\eta_B^2\eta_J}{(\eta_J-\eta_B)(\eta_B\eta_J-\eta_B-\eta_J)}
   (\sigma_A^2+\sigma_B^2) \nonumber \\
  &-&\frac{2\eta_B}{\eta_J-\eta_B}
   +\frac{\eta_B(1-\eta_J \sigma_B^2)}{\eta_B\eta_J-\eta_B-\eta_J}, \\
F &=&
  \eta_J^2\frac{\eta_B+\eta_J-2}{\eta_B-\eta_J}C,\\
G &=& \eta_J^2
   \left(3+\sigma_B^2+\sigma_J^2-C\right)- 
   \frac{2\eta_J(1-\eta_J)D}{\eta_B\eta_J-\eta_B-\eta_J},\\
H &=&
    3-\frac{F}{\eta_B(\eta_B-2)-\eta_J(\eta_J-2)} \nonumber \\
   &+& \frac{G}{\eta_J(\eta_J-2)}
       +\frac{2\eta_J}{\eta_B-\eta_J}E.  \label{eqn:tildeC}
\end{eqnarray}

%**********************************************************
\section{Results and discussion}
\begin{figure}[htbp]
\begin{center}
\includegraphics[width=\gs2\linewidth,keepaspectratio]{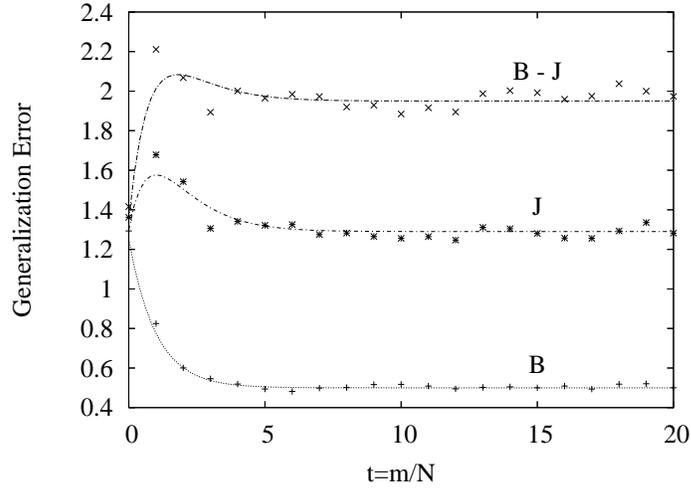}
%-----------------------------------------------
\caption{Generalization errors 
$\epsilon_{Jg}$, $\epsilon_{Bg}$, and $\epsilon_{BJg}$
in the case of $\eta_J=1.2$.
Theory and computer simulation.
Conditions other than $\eta_J$
are $\eta_B=1.0, \sigma_{A}^2=0.2, \sigma_B^2=0.3$,
and $\sigma_J^2=0.4$.}
%-----------------------------------------------
\label{fig:egEB10EJ12VBH02VB03VJ04}
\end{center}
\end{figure}

\begin{figure}[htbp]
\begin{center}
\includegraphics[width=\gs2\linewidth,keepaspectratio]{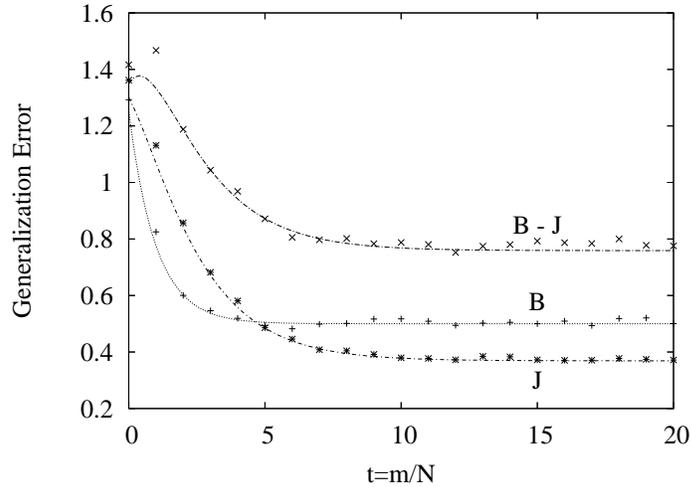}
%-----------------------------------------------
\caption{Generalization errors 
$\epsilon_{Jg}$, $\epsilon_{Bg}$, and $\epsilon_{BJg}$
in the case of $\eta_J=0.3$.
Theory and computer simulation.
Conditions other than $\eta_J$
are $\eta_B=1.0, \sigma_{A}^2=0.2, \sigma_B^2=0.3$,
and $\sigma_J^2=0.4$.}
%-----------------------------------------------
\label{fig:egEB10EJ03VBH02VB03VJ04}
\end{center}
\end{figure}

\begin{figure}[htbp]
\begin{center}
\includegraphics[width=\gs2\linewidth,keepaspectratio]{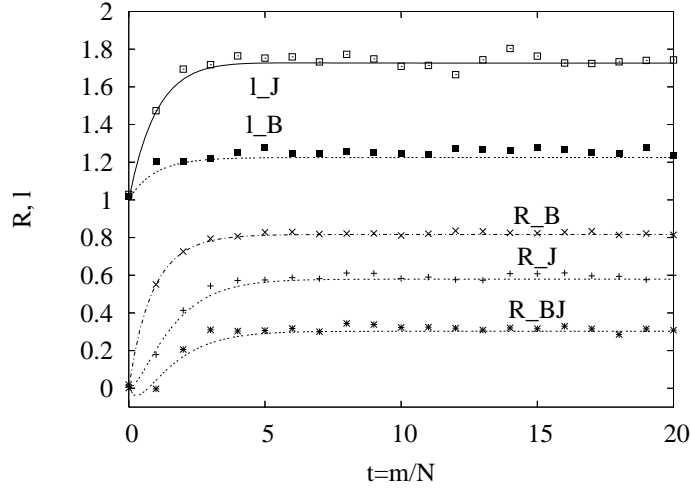}
%-----------------------------------------------
\caption{$R$ and $l$ in the case of $\eta_J=1.2$.
Theory and computer simulation.
Conditions other than $\eta_J$
are $\eta_B=1.0, \sigma_{A}^2=0.2, \sigma_B^2=0.3$,
and $\sigma_J^2=0.4$.}
%-----------------------------------------------
\label{fig:RlEB10EJ12VBH02VB03VJ04}
\end{center}
\end{figure}

\begin{figure}[htbp]
\begin{center}
\includegraphics[width=\gs2\linewidth,keepaspectratio]{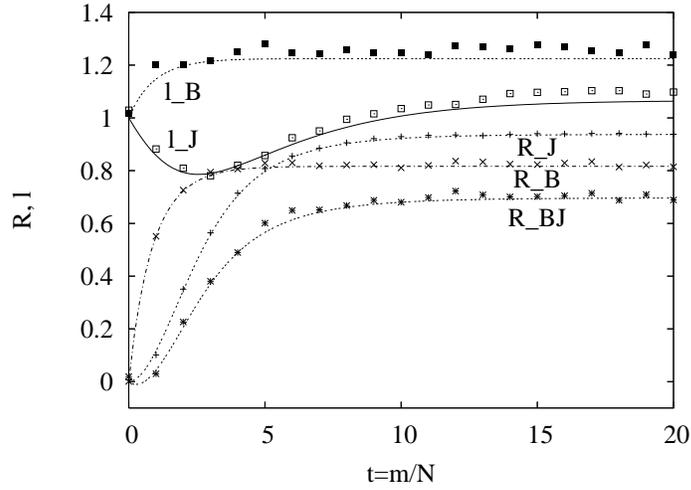}
%-----------------------------------------------
\caption{$R$ and $l$ in the case of $\eta_J=0.3$.
Theory and computer simulation.
Conditions other than $\eta_J$
are $\eta_B=1.0, \sigma_{A}^2=0.2, \sigma_B^2=0.3$,
and $\sigma_J^2=0.4$.}
%-----------------------------------------------
\label{fig:RlEB10EJ03VBH02VB03VJ04}
\end{center}
\end{figure}

%**********************************************************
The dynamical behaviors of the generalization errors
$\epsilon_{Bg}, \epsilon_{Jg}$ and $\epsilon_{BJg}$
have been obtained analytically by solving 
Eqs.(\ref{eqn:hateg}), (\ref{eqn:bareg}), (\ref{eqn:eg}),
%(\ref{eqn:hatrdef})--(\ref{eqn:LJdef})
(\ref{eqn:hatrdef})--(\ref{eqn:rdef})
, and (\ref{eqn:hatr})--(\ref{eqn:tildeC}).
Figures \ref{fig:egEB10EJ12VBH02VB03VJ04} and
\ref{fig:egEB10EJ03VBH02VB03VJ04}
show the analytical results and the corresponding 
simulation results, where $N=10^3$.
In the computer simulations, 
$\epsilon_{Bg}, \epsilon_{Jg}$, and $\epsilon_{BJg}$
have been obtained by
averaging the squared errors for $10^4$ random inputs 
at each time step.
The dynamical behaviors of $R$ and $l$
are shown in Figs.\ref{fig:RlEB10EJ12VBH02VB03VJ04}
and \ref{fig:RlEB10EJ03VBH02VB03VJ04}.
In these figures, 
the curves represent the theoretical results. 
The dots 
%$\times, +, \ast, \Box,\cdots$
represent the simulation results.
Conditions other than $\eta_J$
are common:
$\eta_B=1.0, \sigma_{A}^2=0.2, \sigma_B^2=0.3$,
and $\sigma_J^2=0.4$.
Figures \ref{fig:egEB10EJ12VBH02VB03VJ04} and
\ref{fig:RlEB10EJ12VBH02VB03VJ04}
show the results in the case of $\eta_J=1.2$.
Figures \ref{fig:egEB10EJ03VBH02VB03VJ04} and
\ref{fig:RlEB10EJ03VBH02VB03VJ04}
show the results in the case of $\eta_J=0.3$.

Figure \ref{fig:egEB10EJ12VBH02VB03VJ04}
shows that 
the generalization error $\epsilon_{Jg}$ 
of the student
is always larger than 
the generalization error $\epsilon_{Bg}$ 
of the moving teacher
when the learning rate of student is
relatively large, such as $\eta_J=1.2$.
In addition, 
the mean $\epsilon_{BJg}$ of the error
between the moving teacher and the student 
is still larger than $\epsilon_{Jg}$.
Figure \ref{fig:RlEB10EJ12VBH02VB03VJ04}
shows that the direction cosine $R_J$
between the true teacher and the student
is always smaller than 
the direction cosine $R_B$
between the true teacher and the moving teacher.

On the contrary,
Fig.\ref{fig:egEB10EJ03VBH02VB03VJ04}
shows that when 
the learning rate of the student is
relatively small, that is $\eta_J=0.3$.
Although
the generalization error $\epsilon_{Jg}$
of the student
is larger than 
the generalization error $\epsilon_{Bg}$
of the moving teacher
in the initial stage of learning,
as in the case of $\eta_J=1.2$,
the size relationship is reversed at $t=4.4$,
and after that
$\epsilon_{Jg}$ is smaller than $\epsilon_{Bg}$.
This means the performance of the student becomes higher than
that of the moving teacher.
In regard to the direction cosine,
Fig.\ref{fig:RlEB10EJ03VBH02VB03VJ04} shows that
though the direction cosine $R_J$ between
the true teacher and the student is smaller than
the direction cosine $R_B$
between the true teacher and the moving teacher
in the initial stage of learning,
the size relationship is reversed at $t=5.2$,
and after that,
$R_J$ grows larger than $R_B$.
This means that 
the student gets closer to the true teacher 
than the moving teacher 
in spite of the student only observing the moving teacher.
The reason why the size relationship reverses
at different times in Fig.\ref{fig:egEB10EJ03VBH02VB03VJ04}
and Fig.\ref{fig:RlEB10EJ03VBH02VB03VJ04} is
that the generalization error depends on not only 
the direction cosines $R_B,R_J$, and $R_{BJ}$ 
but also the lengths $l_B$ and $l_J$
as shown in 
Figs.(\ref{eqn:hateg}), (\ref{eqn:bareg}), and (\ref{eqn:eg})
since linear perceptrons are treated and 
the squared error is adopted as an error in this paper.
In any case, these results show that
the student can have higher level of performance than
the moving teacher. It depends on 
the learning rate $\eta_J$ of the student.
This is a very interesting fact.

In addition, both Figs. \ref{fig:RlEB10EJ12VBH02VB03VJ04} and 
\ref{fig:RlEB10EJ03VBH02VB03VJ04} show
that the direction cosine $R_{BJ}$ between 
the moving teacher and the student
takes a negative value in the initial stage of learning.
That is, the angle between the moving teacher and the student
once becomes larger than in the initial condition.
This means that the student is once delayed.
This is also an interesting phenomenon.

\begin{figure}[htbp]
\begin{center}
\includegraphics[width=\gs2\linewidth,keepaspectratio]{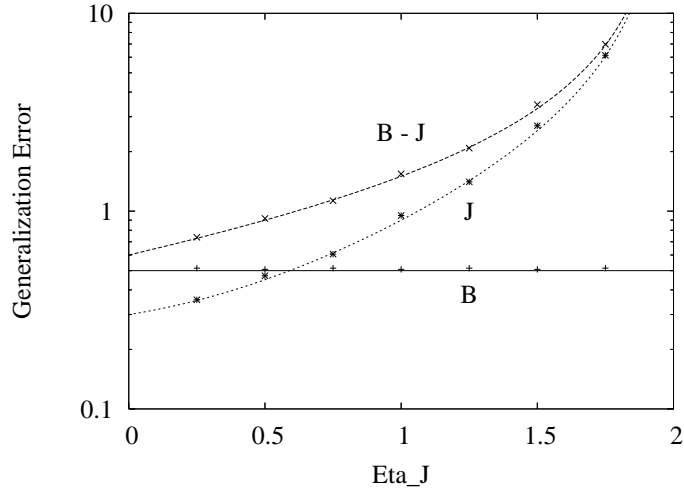}
%-----------------------------------------------
\caption{Steady value of generalization errors
$\epsilon_{Bg}, \epsilon_{Jg}$ and $\epsilon_{BJg}$.
Theory and computer simulation.
Conditions other than $\eta_J$
are $\eta_B=1.0, \sigma_{A}^2=0.2, 
\sigma_B^2=0.3$, and $\sigma_J^2=0.4$.}
%-----------------------------------------------
\label{fig:egEB10VBH02VB03VJ04}
\end{center}
\end{figure}

\begin{figure}[htbp]
\begin{center}
\includegraphics[width=\gs2\linewidth,keepaspectratio]{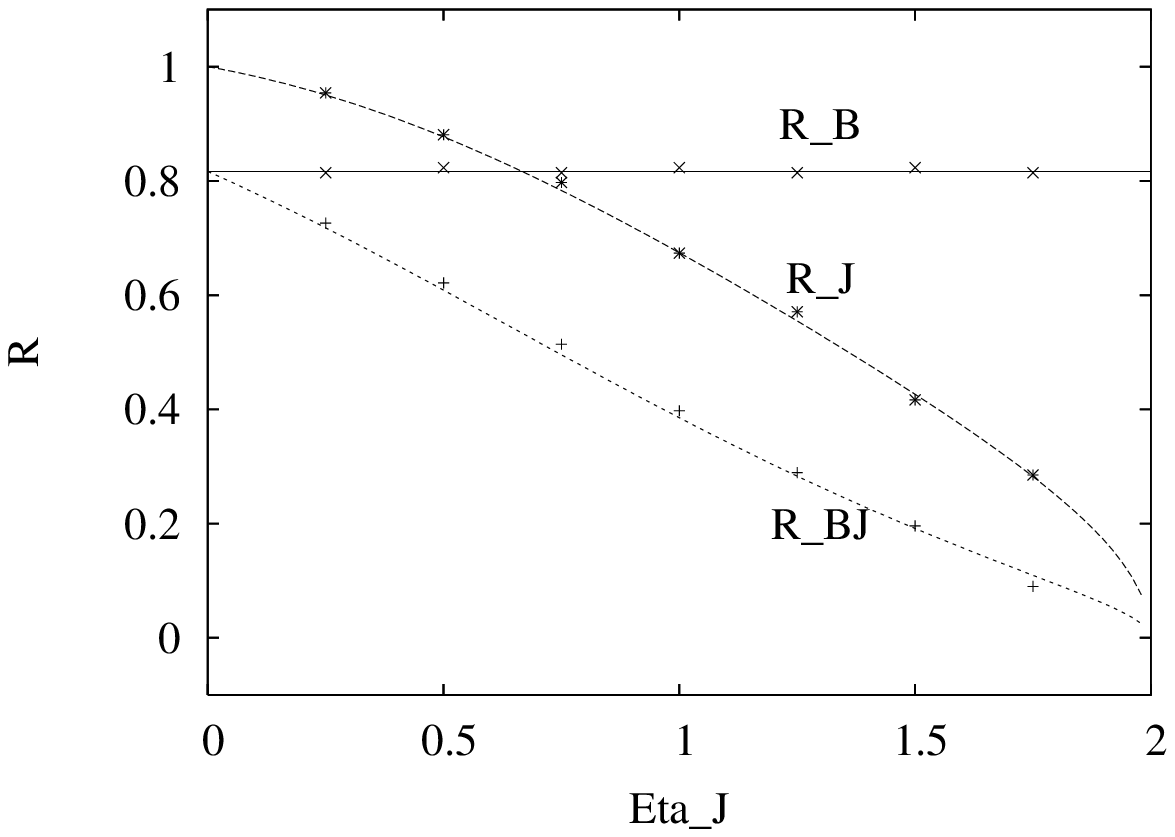}
%-----------------------------------------------
\caption{Steady value of $R$.
Theory and computer simulation.
Conditions other than $\eta_J$
are $\eta_B=1.0, \sigma_{A}^2=0.2, 
\sigma_B^2=0.3$, and $\sigma_J^2=0.4$.}
%-----------------------------------------------
\label{fig:REB10VBH02VB03VJ04}
\end{center}
\end{figure}

\begin{figure}[htbp]
\begin{center}
\includegraphics[width=\gs2\linewidth,keepaspectratio]{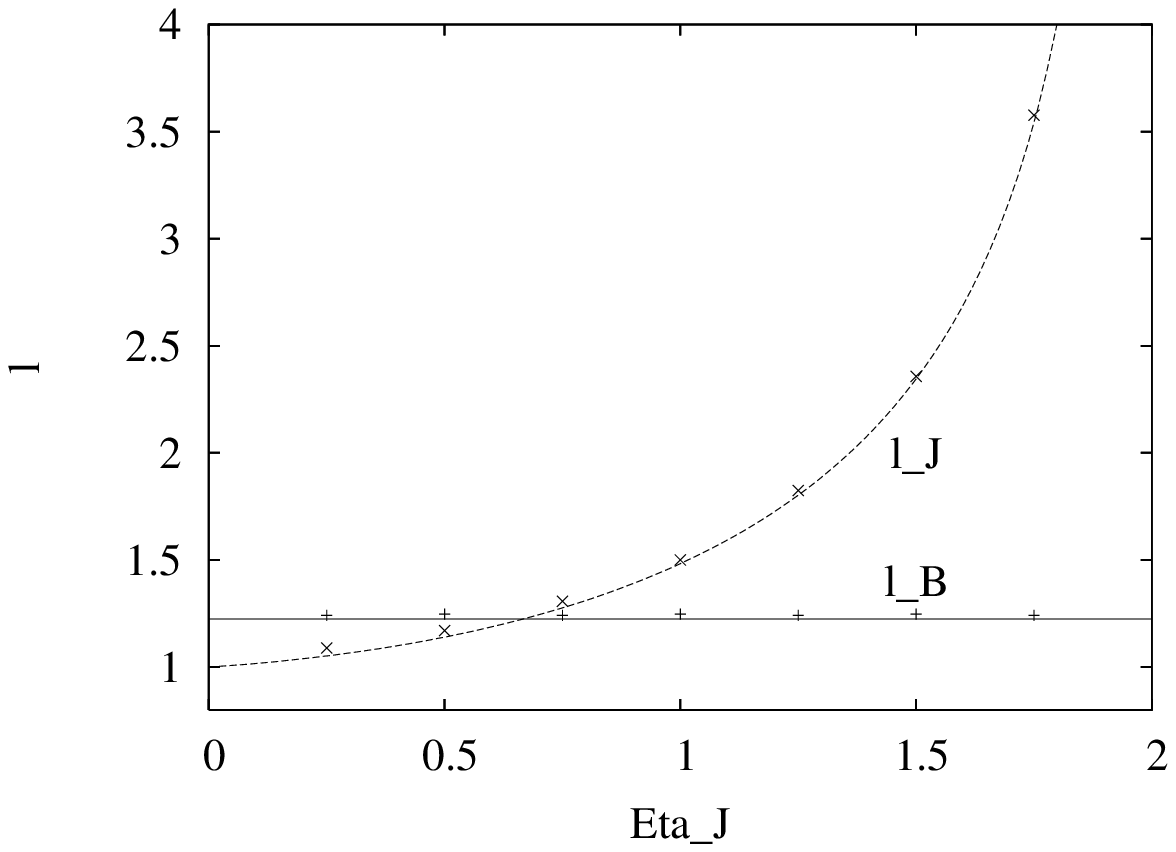}
%-----------------------------------------------
\caption{Steady value of $l$.
Theory and computer simulation.
Conditions other than $\eta_J$
are $\eta_B=1.0, \sigma_{A}^2=0.2, 
\sigma_B^2=0.3$, and $\sigma_J^2=0.4$.}
%-----------------------------------------------
\label{fig:lEB10VBH02VB03VJ04}
\end{center}
\end{figure}

Figures \ref{fig:egEB10EJ12VBH02VB03VJ04} --
\ref{fig:RlEB10EJ03VBH02VB03VJ04}
show that
$\epsilon_{Bg}, \epsilon_{Jg}$, $\epsilon_{BJg}$, 
$R$, and $l$ almost seem to reach a steady state
by $t=20$.
The macroscopic behaviors of $t \rightarrow \infty$
can be understood theoretically 
since the order parameters have been obtained 
analytically.
Focusing on the signs of the powers of the exponential functions
in Eqs.(\ref{eqn:hatr})--(\ref{eqn:LJ}),
we can see that
$\epsilon_{Bg}$ and $\epsilon_{BJg}$ diverge if $0>\eta_B$ or $\eta_B>2$,
and 
$\epsilon_{BJg}$ and $\epsilon_{Jg}$ diverge if $0>\eta_J$ or $\eta_J>2$.
The steady state values of 
$\epsilon_{Bg}, \epsilon_{Jg}$, $\epsilon_{BJg}$, $R$, and $l$
in the case of $0<\eta_B, \eta_J<2$
can be easily obtained by substituting $t\rightarrow \infty$
in Eqs.(\ref{eqn:hatr})--(\ref{eqn:LJ}).
The relationships that are obtained by this operation, 
between the learning rate $\eta_J$ of the student
and 
$\epsilon_{Bg}, \epsilon_{Jg}$, $\epsilon_{BJg}$, 
$R$, and $l$, are shown 
in Figs. \ref{fig:egEB10VBH02VB03VJ04}, \ref{fig:REB10VBH02VB03VJ04},
and \ref{fig:lEB10VBH02VB03VJ04}.
The conditions other than $\eta_J$
are 
$\eta_B=1.0, \sigma_{A}^2=0.2, 
\sigma_B^2=0.3$, and $\sigma_J^2=0.4$
that are the same as Figs. \ref{fig:egEB10EJ12VBH02VB03VJ04}--
\ref{fig:RlEB10EJ03VBH02VB03VJ04}.
The values on $t=50$ are plotted for the simulations.
The values are considered to have already reached a steady state.

These figures show the following:
though the steady generalization error of 
the student
is larger than that of the moving teacher
if $\eta_J$ is larger than 0.58,
the size relationship is reversed if $\eta_J$ is smaller than 0.58.
This means the student has higher level of performance than the moving teacher
when $\eta_J$ is smaller than 0.58.
In regard to the steady $R$ and the steady $l$,
the size relationships are reversed when $\eta_J=0.70$.
In the limit of $\eta_J \rightarrow 0$,
$l_J$ approaches unity,
$R_{BJ}$ approaches $R_B$, and 
$R_J$ approaches unity.
That is, the student $\mbox{\boldmath $J$}$ coincides with
the true teacher $\mbox{\boldmath $A$}$
in both direction and length 
when $\eta_J \rightarrow 0$.
Note that the reason why
the generalization error $\epsilon_{Jg}$ 
of the student isn't zero in 
Fig. \ref{fig:egEB10VBH02VB03VJ04}
is that independent noises are added to the true teacher and the student.
The phase transition in which $R_J$ and $R_{BJ}$ become zero
and $l_J$, $\epsilon_{BJg}$, and $\epsilon_{Jg}$
diverge on $\eta_J=2$ is shown in 
Figs. \ref{fig:egEB10VBH02VB03VJ04}--\ref{fig:lEB10VBH02VB03VJ04}.

%**********************************************************
\section{Conclusion}
%**********************************************************
The generalization errors of 
a model composed of a true teacher,
a moving teacher, and a student 
that are all linear perceptrons with noises
have been obtained analytically
using statistical mechanics.
It has been proven that
the generalization errors of a student
can be smaller than
that of a moving teacher,
even if the student only uses examples 
from the moving teacher.

%**********************************************************
\section*{Acknowledgments}
%**********************************************************
This research was partially supported by the Ministry of Education, 
Culture, Sports, Science, and Technology, Japan, 
with a Grant-in-Aid for Scientific Research
14084212, 14580438, 15500151 and 16500093.

%**********************************************************

\end{document}